\documentclass[a4paper, twocolumn, accepted=2025-12-01]{quantumarticle}
\pdfoutput=1

\usepackage{graphicx}
\usepackage{dcolumn}
\usepackage{bm}
\usepackage{url}
\usepackage[style=numeric-comp,backend=biber,maxnames=1,minnames=1,url=false,doi=true,eprint=false]{biblatex}
\usepackage{hyperref}
\usepackage{physics}
\usepackage{bm}

\usepackage[normalem]{ulem}

\usepackage{color}

\DeclareMathOperator{\HeT}{{}^3\mathrm{He}^\ast}
\DeclareMathOperator{\HeF}{{}^4\mathrm{He}^\ast}

\DeclareMathOperator{\HeM}{\mathrm{He}^\ast}

\newcommand{\tHeT}{$\HeT$}
\newcommand{\tHeF}{$\HeF$}
\newcommand{\tHeM}{$\HeM$}

\newcommand{\thet}{$\HeT$}
\newcommand{\thef}{$\HeF$}

\DeclareMathOperator{\vac}{\ket{\mathrm{vac}}}

\newcommand{\bmk}{{\bm k}}
\newcommand{\bmp}{{\bm p}}
\newcommand{\bmq}{{\bm q}}
\newcommand{\bmr}{{\bm r}}

\newcommand{\doa}{{\downarrow}}
\newcommand{\upa}{{\uparrow}}

\newcommand{\ULA}{\nwarrow}
\newcommand{\URA}{\nearrow}
\newcommand{\DLA}{\swarrow}
\newcommand{\DRA}{\searrow}
\newcommand{\mA}{{\mathcal A}}
\newcommand{\mB}{{\mathcal B}}

\newcommand{\A}{\mathcal A}
\newcommand{\B}{\mathcal B}

\newcommand{\X}{\mathcal X}

\newcommand{\codt}{\cdot}	

\usepackage{soul}
\usepackage[dvipsnames]{xcolor}

\renewcommand{\figurename}{Fig.}

\begin{document}

\title{Proposal for a Bell Test with Entangled Atoms of Different Mass}
\author{X. T. Yan}
\author{S. Kannan}
\author{Y. S. Athreya} 
\author{A. G. Truscott}
\author{S. S. Hodgman}
\email{sean.hodgman@anu.edu.au}
\affiliation{%
	Research School of Physics, Australian National University, Canberra 2601, Australia
}
\date{2025 Nov 11} 

\begin{abstract}
We propose a Bell test experiment using momentum-entangled atom pairs of different masses, specifically metastable helium isotopes \thet and \thef, though the method extends to other atom species.
Entanglement is generated via collisions, after which the quantum states are manipulated using two independent atom interferometers, enabling precise phase control over each species. 
Numerical simulations predict a significant violation of Bell's inequality under realistic conditions.
This proposal opens a new paradigm to study the intersection of quantum mechanics and gravity.
\end{abstract}
\maketitle

\newcommand{\figOne}{
\begin{figure*}
\includegraphics[width=0.99\textwidth]{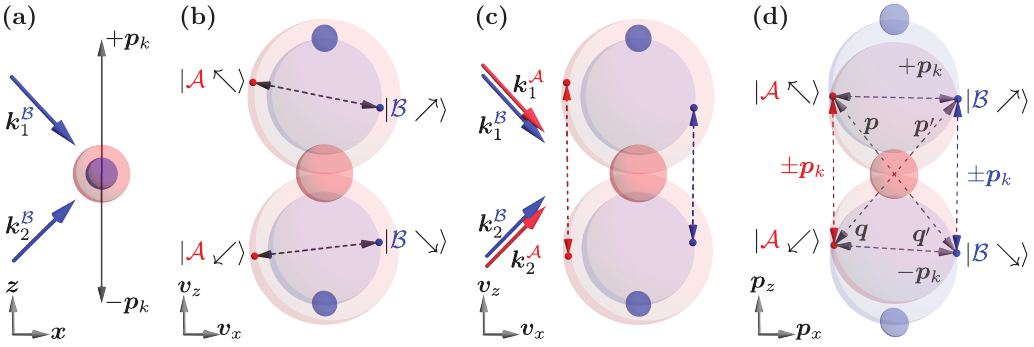}
\caption{
Schematic of our proposed Bell test. 
\textbf{(a)} Initially, we prepare a \thet DFG ($\mA$ - shown in red) and a \thef BEC ($\mB$ - shown in blue) overlapping in a trapping potential (not shown).
    The potential is switched off, and we apply laser beams with wavevectors $k_1^\mB$ and $k_2^\mB$ to induce a Bragg process, transferring \thef ($\mB$) into an equal superposition of two momentum states $\pm \hbar (\bmk_2^\mB-\bmk_1^\mB)=: \pm \bmp_k$. 
    These momentum components travel through the at-rest $\mA$ atoms and cause individual pairs of $\mA$ and $\mB$ atoms to undergo $s$-wave collisions. 
\textbf{(b)} 
    These collisions form two scattering halos in velocity space (faded red and blue spheres) for each of the two initial momenta $\pm \bmp_k$. 
    By energy-momentum conservation, each $\mA$ atom at a particular point on the $\mA$ scattering halo will have a corresponding $\mB$ atom on the diametrically opposite point on the corresponding $\mB$ halo. 
    Two example pairs $\ket{\mA\ULA}\ket{\mB\URA}$ from the $+\bmp_k$ halos and $\ket{\mA\DLA}\ket{\mB\DRA}$ from the $-\bmp_k$ halos are indicated.
\textbf{(c)} 
    The addition of a second pair of Bragg beams resonant with $\mA$ atoms that overlap with the $\mB$ beams $\hat \bmk_{1,2}^\mA=\hat \bmk_{1,2}^\mB$, 
    allows us to couple the upper and lower halo atoms of the same species $\ket{\mA\ULA} \leftrightarrow \ket{\mA\DLA}$ and $\ket{\mB \URA}\leftrightarrow \ket{\mB\DRA}$ independently, 
        which enables the mixing between states for a Bell test. 
\textbf{(d)} 
    Corresponding momentum space distribution of the collision generated states $\ket{\psi}_\mathrm{upper}$ and $\ket{\psi}_\mathrm{lower}$ as described in Eqn \eqref{eq:psi-halo-upper}.  See main text for details.
}
\label{fig:all-in-one}
\end{figure*}
}
\newcommand{\figTwo}{
\begin{figure}
\begin{center}
\includegraphics[width=0.95\columnwidth]{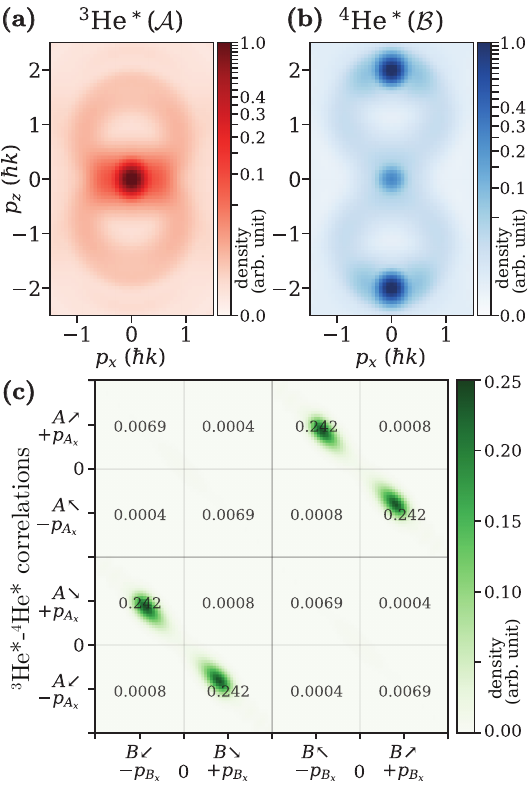}
\end{center}
\caption{ 
Numerical simulations of the momentum density distribution of \textbf{(a)} \thet ($\mA$) and \textbf{(b)} \thef ($\mB$) immediately after the initial Bragg diffraction and collision show the scattering halos generated from $s$-wave collisions between atoms of both species.
\textbf{(c)} The back-to-back correlations present in the dual-species double halo, calculated by slicing the $2D\otimes 2D$ total momentum wavefunction to $1D\times1D$, i.e., $|\braket{p^{\mA}_{z},p^{\mB}_{z}}{\Phi(p^{\mA}_{x},p^{\mA}_{z},p^{\mB}_{x},p^{\mB}_{z})}|^2$. 
    By integrating around the selected momentum states, we can calculate each joint detection probability between different momentum states of interest, shown in the overlaid text. 
    These probabilities are used to calculate the Bell correlation (see Eqn. \eqref{eq:e-corr-def}) of this initial state to be $E=0.94$.
}
\label{fig:all-results}
\end{figure}        
}
\newcommand{\figThree}{
\begin{figure}
\includegraphics[width=\columnwidth]{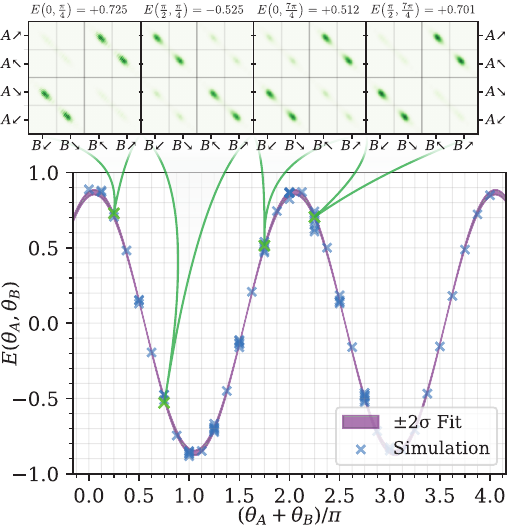} 
\caption{
\textbf{(a)}  
    The Bell correlator $E(\theta_\mA,\theta_\mB)$ for different transfer proportions of $\theta_\mA$ and $\theta_\mB$. 
The simulations show excellent agreement with the quantum prediction of 
    $E=+\cos(\theta_\mA+\theta_\mB)$. 
The four insets show the correlator for the four different phase settings used to calculate the CHSH parameter with a maximal violation. 
For the four values highlighted, the CHSH value exceeds the classical limit of $2$. 
}
\label{fig:results}
\end{figure}
}



\section{Introduction}\label{sec:intro}
One of the most intriguing features of quantum mechanics is the phenomenon of entanglement, 
    where two or more entangled particles exhibit seemingly non-local correlations that defy classical intuition, challenging the concept of local realism. 
    This has led to suggestions that quantum mechanics might be incomplete \cite{einstein1935CanQuantumMechanicalDescription} and that hidden variables could account for the observed correlations to preserve local realism. 
To address this, John Bell formulated a way to experimentally distinguish between the predictions of quantum mechanics and those of local hidden variable theories (LHVTs) through measuring a violation of a so-called Bell inequality \cite{bell1964EinsteinPodolskyRosen}, later made more experimentally accessible by Clauser-Horne-Shimony-Holt (CHSH) \cite{clauser1969ProposedExperimentTest}. 
Historically, Bell tests have predominantly investigated various photonic degrees of freedom \cite{aspect1981ExperimentalTestsRealistic,aspect1982ExperimentalTestBells,tomasin2017HighvisibilityTimebinEntanglement,heaney2009BellinequalityTestSpatialmode,chen2022ImagingSinglePhotonOrbitalAngularMomentum,rarity1990ExperimentalViolationBells,rarity1990TwocolorPhotonsNonlocality}. 
While lesser in number, Bell tests using atoms and electrons have so far primarily explored the internal degrees of freedom such as spin \cite{rowe2001ExperimentalViolationBells,hensen2015LoopholefreeBellInequality,shin2019BellCorrelationsSpatially,schmied2016BellCorrelationsBoseEinstein}. 
While these experiments have provided strong evidence against local hidden variable theories, there remains a lack of experimental evidence for Bell tests involving the external degrees of freedom (momentum) of massive particles.

Attempts at atomic momentum entanglement for Bell tests have faced significant experimental challenges \cite{pezze2018QuantumMetrologyNonclassical}.
Dussarrat \textit{et al.} \cite{dussarrat2017TwoParticleFourModeInterferometer} experimentally interfered different modes of two particles and ruled out the possibility of a mixed state. 
Fadel \textit{et al.} \cite{fadel2018SpatialEntanglementPatterns} 
    and Lange \textit{et al.} \cite{lange2018EntanglementTwoSpatially} showed spatial entanglement in cold atom experiments.
Shin \textit{et al.} \cite{shin2019BellCorrelationsSpatially} demonstrated Bell correlations between separated atomic pairs, using momentum as their particle label and atomic spin states to show entanglement in spin. 
Other attempts to produce entanglement in atomic momentum  \cite{tomkovic2011SingleSpontaneousPhoton,slodicka2013AtomAtomEntanglementSinglePhoton,dussarrat2017TwoParticleFourModeInterferometer,anders2021MomentumEntanglementAtom}, 
while successfully showing forms of non-classical correlations, possessed insufficient coherence to perform a full Bell test.
Our recent attempt \cite{thomas2022MatterwaveRarityTapsterInterferometer,athreya2025bellcorrelationsmomentumentangledpairs} 
used mixing between different momentum states in two separate $s$-wave scattering halos of metastable helium-4 (\tHeF) atoms to measure Bell correlations.  However, due to the lack of independent phase control, a full CHSH Bell test \cite{lewis-swan2015ProposalMotionalstateBell} was not possible, meaning only a limited class of LHVTs could be excluded.

In this paper, we propose an experiment to test a Bell inequality using momentum-entangled atoms of different mass, specifically two isotopes of \tHeM, although the scheme would also work with other species. 
By inducing controlled collisions between these isotopes and utilising species-specific momentum-transfer pulses for independent control, 
    we generate and manipulate entangled pairs suitable for a complete CHSH Bell test. 
We present theoretical modeling of the entangled state generation, 
    followed by numerical simulations that demonstrate a violation of the Clauser-Horne-Shimony-Holt inequality under realistic experimental conditions.

\section{Mass and Momentum Entanglement}\label{sec: mass-entanglement}
The system considered here features two objects of different mass in a superposition of external degrees of freedom such as momentum, which appears to be a relatively underexplored quantum resource.  We term such a state mass entangled.  Such a system has the potential to shed insight into a range of interesting physics at the intersection of quantum mechanics and gravity \cite{marletto2017GravitationallyInducedEntanglement,geiger2018ProposalQuantumTest,carney2019TabletopExperimentsQuantum}, by testing how the motion of the different masses in the superposition is affected by a gravitational field.  One of the simplest such states features a superposition of two masses $\mA$ and $\mB$, in two different spatial locations $\leftarrow$ and $\rightarrow$, written as
\begin{align}
{\ket{\psi} = \ket{\mA,\leftarrow}\ket{\mB,\rightarrow} + \ket{\mB,\leftarrow}\ket{\mA,\rightarrow}}.
\label{eq:mass-ent}
\end{align}
This state could be generated experimentally, e.g. through s-wave collisions between ensembles of  colliding $\mA$ and $\mB$ atoms, with $\mA$ initially in momentum state $\leftarrow$ and $\mB$ initially in momentum state $\rightarrow$, similar to a previous experiment with different spin states \cite{shin2019BellCorrelationsSpatially}.

However, while the state in Eqn. \ref{eq:mass-ent} is interesting, it appears to be practically impossible to prove that it is entangled, as there is no feasible mechanism to mix the $\mA$ and $\mB$ atoms and it is challenging to mix the momentum modes in an identical manner for the two different atoms.  A way to proceed is to introduce two additional momentum modes, such that we now consider entanglement between the two pairs of momentum mode pairs ($\ULA$ and $\URA$), and ($\DLA$ and $\DRA$), as well as the masses as above.  The full state now becomes
\begin{align}
&
\ket{\psi} = \ket{\mA,\ULA}\ket{\mB,\URA}+\ket{\mB,\ULA}\ket{\mA,\URA} 
\notag
\\
&
+ \ket{\mA,\DLA}\ket{\mB,\DRA} + \ket{\mB,\DLA}\ket{\mA,\DRA},
\label{eq:8mode-ent}
\end{align}
which represents a hyperentangled state \cite{DengHyperentanglement2017} in momentum and mass.  As with the state described in Eqn. \ref{eq:mass-ent}, such a state could be generated using proven experimental techniques, by extending 4-mode momentum entanglement to two atomic species \cite{athreya2025bellcorrelationsmomentumentangledpairs}.  However, once again any full test of this entanglement would require a way to mix the masses, which is unfeasible.  

An alternative way to demonstrate entanglement in such a state is to truncate this state by freezing out the mass entanglement and only considering half of the cases:
\begin{align}
    \ket{\psi} = \ket{\mA \ULA}\ket{\mB \URA} + \ket{\mA \DLA}\ket{\mB \DRA}. 
    \label{eq:34momentum-ent}
\end{align}
Such a state can be used to perform a full CHSH Bell test with massive particles entangled in an external degree for freedom, for the first time, by mixing between the momentum modes of each atomic species independently.  The following sections consider how to conduct a CHSH test by mixing the momentum modes of the state in Eqn. \ref{eq:34momentum-ent}.

\figOne

\section{Proposed Experimental Setup}\label{sec: setup}
\paragraph*{Entangled state generation.}
We start by creating an entangled state through collisions between two atomic species, in our case, the fermionic and bosonic isotopes of metastable helium \cite{thomas2023ProductionHighlyDegenerate}, denoted as $\mA$ (\thet) and $\mB$ (\thef) respectively, initially in a harmonic trap. 
After switching off the trap, we transfer the $\mB$ atoms into a coherent superposition of momentum states moving upwards ($+\bmp_k$) and downwards ($-\bmp_k$), while the $\mA$ atoms remain at rest (see \figurename~\ref{fig:all-in-one}.(a)). 
The transfer is achieved via a two-photon Bragg transition, transferring momentum $\pm\bmp_k$ to the atoms, 
    where $\bmp_k = \hbar(\bmk_2^\mB-\bmk_1^\mB)$ and $\bmk_{1,2}^\mB$ are the two beam wavevectors associated with the laser beams that induce the two-photon transition of $\mB$ 
    \cite{giltner1995TheoreticalExperimentalStudy, thomas2022MatterwaveRarityTapsterInterferometer}. 
This momentum transfer leads to $s$-wave collisions between individual \thet and \thef atoms, 
    with collisions occurring between pairs with centre-of-mass  momenta either $+\bmp_k$ and 0 or $-\bmp_k$ and 0. 
Each of these collisions will produce two scattering halos \cite{perrin2007ObservationAtomPairs,hodgman2017SolvingQuantumManyBody,kannan2024MeasurementSwaveScattering}, 
    such that for every \thet atom detected in one halo, a \thef atom will be found in the diametrically opposite point in the other halo of the pair (see \figurename~\ref{fig:all-in-one}.(b)). 
Notably, \thef-\thef scatterings are absent in this setup, as there are no \thef atoms left in the zero centre-of-mass momentum states. While scattering between $+\bmp_k$ and $-\bmp_k$ for \thef atoms is possible, as has previously been observed \cite{Khakimov2016}, atoms in the resulting much larger halo will land outside our analysis region and hence not contribute \cite{hodgman2017SolvingQuantumManyBody,Khakimov2016}.

The effective Hamiltonian $\hat H_\mathrm{FWM}$ describing the four-wave mixing (FWM) process responsible for the pair production 
    comes from extending the previous theoretical framework of FWM for pure bosonic atoms 
    \cite{klimov2009GroupTheoreticalApproachQuantum,braunstein2005QuantumInformationContinuous,rugway2011CorrelationsAmplifiedFourWave,thomas2022MatterwaveRarityTapsterInterferometer} and pure fermions \cite{khanna2007MaximumEntanglementSqueezed}, 
    to a Bose-Fermi (\thef-\thet) mixture system.
This gives: 
\begin{align}
\hat H_\mathrm{FWM} = \sum_{\bmp} \hbar \zeta \big( \hat b_{\bmp}^\dagger \hat f_{\bmp'}^\dagger + \hat b_{\bmp} \hat f_{\bmp'} \big), 
\label{eq:fwm-ham}
\end{align}
where $\hat b_{\bmp}^\dagger$ and $\hat f_{\bmp'}^\dagger$ are the bosonic and fermionic creation operators for the \thef and \thet atoms, respectively.
$\zeta$ is the coupling strength of this interaction depending on the scattering length and density.
The summation over momenta $\bmp$ accounts for the \thef atoms with momentum $\bmp$ and \thet atoms with momentum $\bmp'$ that satisfy both momentum conservation $\bmp+\bmp' = +\bmp_k$ and energy conservation. 
Consequently, only specific outgoing momenta $p$ and $p'$ are allowed, resulting in the formation of spherical halos in velocity and momentum space as depicted in \figurename~\ref{fig:all-in-one}.(b) and \figurename~\ref{fig:all-in-one}.(d) respectively.

An initial vacuum state $\vac$ then evolves under the unitary evolution operator to obtain $U(t)$ as
\begin{align}
    \hat U(t) \vac &= e^{-\frac{i}{\hbar} \hat H t} \vac
\\
&=
    \vac + 
    i\zeta t \sum_{\bmp} \hat b_{\bmp}^\dagger \hat f_{\bmp'}^\dagger \vac
    + \cdots
\label{eq:ut-vac}
\end{align}
where higher-order terms involving the creation of multiple pairs into the same $(\bmp,\bmp')$ are not possible due to the antisymmetry of fermions (Pauli exclusion) \cite{khanna2007MaximumEntanglementSqueezed}, 
    leaving only the vacuum state or one pair of excitations in the same $p$.
The state after FWM can therefore be approximated as a two-mode squeezed vacuum state for each scattering direction around the halo.

For the upper halo (momentum $+\bmp_k$), the state is
\begin{align}
    \ket{\psi}_\mathrm{upper}
&\simeq
    \sum_{\bmp} \left(
    \ket{0}^\mA_{\bmp} \ket{0}^\mB_{\bmp'} + 
    \tilde\zeta \ket{1}^\mA_{\bmp} \ket{1}^\mB_{\bmp'}
    \right),
\label{eq:psi-halo-upper}
\end{align}
where $\tilde{\zeta} = i\zeta t$ and $\ket{n}^{\mA, \mB}_{\bmp, \bmp'}$ denote an $n$-atom Fock state of \thet and \thef atoms with momenta $\bmp$ and $\bmp'$, respectively. 
The lower halo will be the same but replacing $\bmp, \bmp'$ with $\bmq, \bmq'$,
with $\bmq + \bmq' = -\bmp_k$.

The total state is then the tensor product of the upper and lower halo states: 
\begin{align}
    \ket{\Psi}
&=  
    \ket{\psi}_\mathrm{upper}\ket{\psi}_\mathrm{lower}
\notag
\\
&= \phantom{\otimes}
    \sum_{\bmp} \left(
    \ket{0}^\mA_{\bmp} \ket{0}^\mB_{\bmp'} + 
    \tilde\zeta \ket{1}^\mA_{\bmp} \ket{1}^\mB_{\bmp'}
    \right)
\notag
\\
& \phantom{=} \  \otimes
    \sum_{\bmq} \left(
    \ket{0}^\mA_{\bmq} \ket{0}^\mB_{\bmq'} + 
    \tilde\zeta \ket{1}^\mA_{\bmq} \ket{1}^\mB_{\bmq'}
    \right)
\notag
\\
&=
    \sum_{\bmp}\sum_{\bmq}
    \ket{0}^\mA_{\bmp} \ket{0}^\mB_{\bmp'} \ket{0}^\mA_{\bmq} \ket{0}^\mB_{\bmq'}
\notag
\\
&\phantom{.}
    +
    \tilde\zeta \ket{1}^\mA_{\bmp} \ket{1}^\mB_{\bmp'} \ket{0}^\mA_{\bmq} \ket{0}^\mB_{\bmq'}
    +
    \tilde\zeta \ket{0}^\mA_{\bmp} \ket{0}^\mB_{\bmp'} \ket{1}^\mA_{\bmq} \ket{1}^\mB_{\bmq'}
\label{eq:Psi-total-state-pair}
\\
&\phantom{.}
    +
    \tilde\zeta^2 \ket{1}^\mA_{\bmp} \ket{1}^\mB_{\bmp'} \ket{1}^\mA_{\bmq} \ket{1}^\mB_{\bmq'}. 
\label{eq:Psi-total-state-HOM}
\end{align}
These terms represent the vacuum state, the single pair production in the upper and lower halo, and the two-pair production, respectively.

Since $\zeta \ll 1$ \cite{thomas2022MatterwaveRarityTapsterInterferometer,lewis-swan2015ProposalMotionalstateBell}, the two-pair term \eqref{eq:Psi-total-state-HOM} is negligible compared to the single-pair terms \eqref{eq:Psi-total-state-pair}.
We can also post-select events where only one pair of atoms is produced and focus on the superposition of the second and third terms. 
As shown in \figurename~\ref{fig:all-in-one}.(d), we select $\bmp$, $\bmq$ momenta satisfying $\bmp-\bmq=\bmp_k$. 
Notably, this geometry guarantees that the corresponding momenta $\bmp'$ and $\bmq'$ also satisfy $\bmp'-\bmq'=\bmp_k$, 
thus achieving a consistent vertical momentum transfer across selected pairs, realising a momentum mirror and beam-splitter. 
This is required due to the different momenta of the \thet and \thef components of each of the two halos. 
For notational simplicity, we define:
\begin{align}
\ket{\mA,\ULA} &\equiv \ket{1}^\mA_{\bmp}\ket{0}_{\bmq}^\mA \ , &
\ket{\mA,\DLA} &\equiv \ket{0}^\mA_{\bmp}\ket{1}_{\bmq}^\mA \ , 
\notag
\\
\ket{\mB,\URA} &\equiv \ket{1}^\mB_{\bmp'}\ket{0}_{\bmq'}^\mB \ , &
\ket{\mB,\DRA} &\equiv \ket{0}^\mB_{\bmp'}\ket{1}_{\bmq'}^\mB \ .
\label{eq:mom-states-relabel}
\end{align}

Thus, the entangled state of interest becomes: 
\begin{align}
    \ket{\psi} = \ket{\mA \ULA}\ket{\mB \URA} + \ket{\mA \DLA}\ket{\mB \DRA}, 
    \label{eq:initial-state}
\end{align}
where the arrows denote their momentum directions, which is the same as Eqn. \ref{eq:34momentum-ent}.

Although the collision excites many pairs in various directions, we can concentrate only on pairs along a specific direction for clarity. 
This is possible as the Pauli exclusion principle ensures that 
    the pair creation \eqref{eq:ut-vac} can only generate a maximum of one pair in any particular momentum state.
The cylindrical symmetry of the halos will allow this analysis to then be extended to multiple independent sets of 
    pairs in different directions, 
increasing the experimental data acquisition rate.

\paragraph*{Independent Bragg Pulses.}
Coupling between upper and lower halo states is achieved with separate Bragg laser pulses, tuned in phase and amplitude to function as atomic momentum mirrors and beam-splitters.
These utilise the distinct transition frequencies of 
    the $2^3S_1 \to 2^3P_0$ states of 
    \thet (276.7322 THz) and \thef (276.6986 THz) 
	\cite{thomas2023ProductionHighlyDegenerate,mcnamara2006DegenerateBoseFermiMixture}.
The laser frequencies are adjusted to match the atomic resonances of one isotope while being sufficiently detuned from the other 
    \cite{thomas2023ProductionHighlyDegenerate,thomas2024BodyAntibunchingDegenerate}, 
    allowing independent coupling of each species. 
The Bragg pulse induces momentum transfer through an effective unitary transformation given by 
	\cite{thomas2022MatterwaveRarityTapsterInterferometer}: 
\begin{align}
    \ket{\X , \upa} \xrightarrow{\hat U_{\varphi,\theta}^\X}&
            \cos{\frac{\theta_{\X}}{2}} \ket{\X , \upa} 
        -   ie^{i\varphi_{\X}}\sin{\frac{\theta_{\X}}{2}} \ket{\X , \doa}
    \notag
    \\
    \ket{\X, \doa} \xrightarrow{\hat U_{\varphi,\theta}^\X}&
        -   ie^{-i\varphi_{\X}}\sin{\frac{\theta_{\X}}{2}} \ket{\X , \upa}
        +   \cos{\frac{\theta_{\X}}{2}} \ket{\X , \doa}, 
    \label{eq:beam-splitter}
\end{align}
where $\mathcal X \in \{ \mA,\mB \}$
    and $\uparrow$      represents $\ULA$ for $\mA$ or $\URA$ for $\mB$
    and $\downarrow$    represents $\DLA$ for $\mA$ or $\DRA$ for $\mB$. 
$\theta_\X = \Omega_\X t_\X$, with $\Omega_\X$ the Rabi frequency and $t_\X$ the laser pulse duration,
so that $\theta_\X = \pi$ represents a mirror and
$\theta_\X = \pi/2$ is a beam splitter.  
$\varphi_\X $ is the relative laser phase between the $k_1,k_2$ beams and sets the phases for the Bell test.

\paragraph*{The Bell Test.}
A key requirement for performing a Bell test using momentum-entangled atoms is the ability to mix different momentum states coherently \cite{lewis-swan2015ProposalMotionalstateBell}. 
However, atoms in distinct momentum states propagate along separate trajectories and become spatially separated over time. 
This spatial separation introduces an additional degree of freedom, position, making the states distinguishable. 
To allow effective mixing and interference of the momentum states, we need to bring them back into spatial overlap. 
We achieve this by applying mirror pulses ($\hat U_{0,\pi}^\mA \hat U_{0,\pi}^\mB$) to both species 
at $t_1$ after the entangled state is created, reversing the momentum of each atom. 
This causes the momentum states to converge and overlap in space at a later time $t_2$. 
At the moment of overlap, we perform the beam-splitter operation ($\hat U_{\varphi_\mA,\frac{\pi}{2}}^\mA \hat U_{\varphi_\mB,\frac{\pi}{2}}^\mB$) at $t_2$ with variable phases ($\varphi_\mA$ and $\varphi_\mB$) to perform the Bell test. 

After applying the mirror ($\pi$) pulse and a 50:50 beam-splitter ($\frac{\pi}{2}$) pulse with phases $\varphi_\mA$ and $\varphi_\mB$ to each species, the state evolves to 
\begin{align}
    \ket{\psi_{t_2}} = &
        \hat U_{\varphi_\mA, \frac{\pi}{2}}^\mA \hat U_{\varphi_\mB, \frac{\pi}{2}}^\mB \ket{\psi_{t_1}}
    \\
=& \textstyle \frac{1}{2\sqrt2}\left(
        -1 + e^{-i\varphi_\mA-i\varphi_\mB}
    \right) \ket{\mA\ULA}\ket{\mB\URA}
    \label{eq:psi-t2-uu}
\\
&\textstyle +\frac{i}{2\sqrt2}\left(
        e^{-i\varphi_\mA} + e^{i\varphi_\mB}
    \right) \ket{\mA\ULA}\ket{\mB\DRA}
    \label{eq:psi-t2-ud}
\\
&\textstyle +\frac{i}{2\sqrt2}\left(
        e^{i\varphi_\mA} + e^{-i\varphi_\mB}
    \right) \ket{\mA\DLA}\ket{\mB\URA}
    \label{eq:psi-t2-du}
\\
&\textstyle +\frac{1}{2\sqrt2}\left(
        -1 + e^{i\varphi_\mA+i\varphi_\mB}
    \right) \ket{\mA\DLA}\ket{\mB\DRA}.
    \label{eq:psi-t2-dd}
\end{align}
Thus four joint momentum states are key to the Bell inequality. 
The probabilities for detecting atoms in each of the states $\circ,\bullet $ are 
\begin{align}
P_{ \circ \bullet} =& \big| \braket{\ \circ\ \bullet\ }{\psi_{t_2}}  \big|^2
= \frac{1\pm \cos(\varphi_\mA+\varphi_\mB)}{4}, 
\label{eq:prob-corr}
\end{align}
with the $-$ case for $\bra{\circ\bullet} = \bra{\mA\ULA}\bra{ \mB\URA}, \bra{\mA\DLA}\bra{\mB\DRA}$ and 
the $+$ case for $\bra{\circ\bullet} = \bra{ \mA\ULA \mB\DRA} , \bra{\mA\DLA\mB\URA}$.
This probability $P_{ \circ \bullet}$ then corresponds to the experimentally measurable \cite{thomas2022MatterwaveRarityTapsterInterferometer,shin2019BellCorrelationsSpatially} second-order correlation function defined as \cite{hodgman2017SolvingQuantumManyBody}
\begin{align}
    g_{\circ \bullet}^{(2)} = \frac{\int d\phi\ \langle \colon  {\hat n}_{\circ\phi}^\mA \ \  {\hat n}_{\bullet\phi}^\mB  \colon\rangle }
    {   \int d\phi\ \langle \ {\hat n}_{\circ\phi}^\mA \ \rangle 
        \int d\phi\ \langle \ {\hat n}_{\bullet\phi}^\mB \  \rangle 
    },
\end{align}
where $\hat n = \hat a ^\dagger \hat a $ is the number density, and $\circ,\bullet$ denote the momentum states of species $\mA$ and $\mB$. 
Due to the cylindrical symmetry of the system, the correlation can be integrated over all azimuthal angles $\phi$.

For a Bell test 
    \cite{bell1964EinsteinPodolskyRosen,clauser1969ProposedExperimentTest,shin2019BellCorrelationsSpatially,thomas2022MatterwaveRarityTapsterInterferometer}, 
    we aim to measure the Bell correlation $E$ between the momentum of the two species. 
$E$ is defined as the expectation value of the product of the momentum observables of species $\mA$ and $\mB$: 
$   
    E(\varphi_\mA, \varphi_\mB) = \expval{\hat{\bm{p}}_\mA \ \hat{\bm{p}}_\mB}, 
$
where $\hat{\bm{p}}_\X$ is a two-mode momentum operator of species $\X$. These operators have normalised eigenvalues of $\pm 1$, corresponding to the upwards and downwards momentum states of each species. 
$E$ can then be calculated from the second-order correlation function $g^{(2)}_{\circ\bullet}$ as
\begin{align}
E(\varphi_\mA, \varphi_\mB) 
&= g_{\ULA\URA}^{(2)} + g_{\DLA\DRA}^{(2)} - g_{\ULA\DRA}^{(2)} - g_{\DLA\URA}^{(2)}
\label{eq:e-corr-def}
\\ 
&=  \phantom{+}\textstyle 
    \frac{1-\cos(\varphi_\mA  + \varphi_\mB)}{2} \cos(\theta_\mA-\theta_\mB)
\\
&\phantom{=}\textstyle
    + \frac{1+\cos(\varphi_\mA + \varphi_\mB)}{2} \cos(\theta_\mA+\theta_\mB)
\notag
\\
&= \phantom{-} \cos(\theta_\mA + \theta_\mB) \ \text{ for $\textstyle \varphi_\mA=\varphi_\mB = 0$}
\label{eq:corrEpCosTheta}
\end{align}
for some particular mixing pulse phase settings $\varphi_\mA$ and $\varphi_\mB$.
To test Bell's inequality, we compute the CHSH parameter $\mathcal S$ \cite{clauser1969ProposedExperimentTest,aspect1975ProposedExperimentTest} with four different correlators, 
\begin{align}
    \mathcal S 
    =& \ 
    E(\varphi_\A,\varphi_\B) - E(\varphi_\A,\varphi_\B') \notag \\ 
    &+ E(\varphi_\A',\varphi_\B) + E(\varphi_\A',\varphi_\B')
\label{eq:chsh-param-def}
\end{align}
with each term representing experiments with those specific phase settings. 
For instance, choosing phases 
\begin{align}
    \varphi_\mA + \varphi_\mB  &= \textstyle\frac{\pi}{4}, &
    \varphi_\mA + \varphi_\mB' &= \textstyle\frac{3\pi}{4}, 
\notag
\\  
    \varphi_\mA' + \varphi_\mB &= \textstyle\frac{7\pi}{4}, &
    \varphi_\mA' + \varphi_\mB' &= \textstyle\frac{9\pi}{4}, 
\end{align}
each term in \eqref{eq:chsh-param-def} evaluates to $+1/\sqrt2$, and $\mathcal S = 2\sqrt 2$, exceeding the local hidden variable theory limit of $2$ \cite{clauser1969ProposedExperimentTest}.

\section{Simulations} \label{sec: simulations}
\paragraph{The Split-Step Operator Method.}
We developed a numerical simulation to model the full experimental sequence, including the Bragg pulses, collisions, and beam-splitter operations.
This employs the split-step operator method \cite{james2019MassivelyParallelSplitstep} to solve the time-dependent Schr\"odinger equation for the two-particle wavefunction of \thet and \thef. 
The method alternates between applying the kinetic and potential energy operators in small timesteps, allowing efficient and accurate evolution of the wavefunction in time.
Chosen for its ability to explicitly compute the entangled state of the two species, this approach is also versatile and can be adapted to other studies involving entanglement in continuous variable systems. 

The two-particle wavefunction is discretised on a $2D\otimes 2D $ grid to balance computational efficiency while retaining the scattering halo geometry. 
The wavefunction is expressed as
\begin{align}
    \ket{\psi(\bmr,t)} = \ket{\psi(x_3,z_3,x_4,z_4,t)}
\end{align}
where $x_3,z_3$ and $x_4,z_4$ denote the spatial coordinates of the \thet and \thef atoms respectively.

\figTwo

The time evolution of the wavefunction is governed by 
\begin{align}
    \ket{\psi(\bmr,t + dt )} 
    &= e^{-\frac{i}{\hbar} (\hat H_p + \hat H_r) dt} \ket{\psi(\bmr,t)} 
\end{align}
where the full Hamiltonian can be split into momentum and position dependences $\hat H = \hat H_p + \hat H_r$. 
Here $\hat H_p = \frac{p^2}{2m}$ represents the kinetic term and $\hat H_r$ includes the $s$-wave scattering and Bragg pulse potentials. 
Using the split-step operator method, also known as the Zassenhaus or reverse Campbell-Hausdorff expansion \cite{james2019MassivelyParallelSplitstep,magnus1954ExponentialSolutionDifferential,casas2012EfficientComputationZassenhaus}, the evolution over a time step $dt$ is approximated as
\begin{align}
    \ket{\psi(\bmr,t + dt )} 
    =& \ 
    {e^{-\frac{i}{\hbar} \hat H_r \frac{dt}{2}   }}
    {e^{-\frac{i}{\hbar} \hat H_p dt }   }
    {e^{-\frac{i}{\hbar} \hat H_r \frac{dt}{2}}  }
    \ket{\psi(\bmr,t)} \notag \\ 
    &+ \mathcal O(dt^3).
    \label{eq:split-step}
\end{align}
For our case, we selected a time step of $0.1\mu$s and grid size of 121 points per dimension, providing a spatial resolution of $0.5\mu$m and momentum resolution of $0.066\hbar k$, where $k = |\bmk_1-\bmk_2|$ is the wavevector of the Bragg beams.
This configuration is a balance between accuracy, with $\mathcal O(dt^3) $ error, and computational feasibility, requiring $\sim 10$ hours runtime on a standard desktop computer to simulate the full experimental duration of $\sim 1.2$ms, with a memory footprint of $\sim 4$GB per wavefunction array.

\paragraph{Simulation of the full experiment.}
Building on the simulation method described above, we model the full experiment sequence under realistic experimental conditions. 
The parameters include the laser pulse durations ($50\mu\mathrm{s}$ and $66\mu\mathrm{s}$ for \thet and \thef respectively), intensities ($\sim 0.1 \mathrm{mW}/\mathrm{mm}^2$), as well as the atomic scattering properties $(a_{34}=29\mathrm{nm})$ \cite{kannan2024MeasurementSwaveScattering,mcnamara2006DegenerateBoseFermiMixture,hirsch2021ClosecoupledModelFeshbach}.

Following the scheme described earlier (see \figurename~\ref{fig:all-in-one}), we initialise overlapping degenerate gases of \thet and \thef atoms in the groundstate of the harmonic trapping potential. 
To simulate the subsequent mirror and beam-splitter, we apply an effective Bragg pulse potential given by $\hat V_B = V_B(t)\cos(2\bmk \cdot \bm{x} - \delta t - \varphi)$, where $V_B(t)$ is the pulse envelope, $\bmk$ is the wavevector, $\delta$ is the detuning, and $\varphi$ is the phase \cite{thomas2022MatterwaveRarityTapsterInterferometer}.

In \figurename~\ref{fig:all-results}.(a) and (b), we show the momentum density distribution of \thet and \thef, respectively, immediately after the initial Bragg diffraction and collision, highlighting the resultant $s$-wave scattering halos.
By analysing slices of the wavefunction, 
    we confirm that strong correlations are present in the initial state, as shown by the correlation amplitude plot in \figurename~\ref{fig:all-results}.(c).

\paragraph{Observation of Bell correlations.}
We then add mirror and mixing pulses to simulate the full Bell test sequence with varying mixing pulse transfer fraction $\theta_\mA$ and $\theta_\mB$ to produce the full correlation plot shown in \figurename~\ref{fig:results}. 
The correlation function $E(\theta_\mA,\theta_\mB)$ closely follows the quantum mechanical prediction of $\cos(\theta_\mA+\theta_\mB)$ from Eqn.~\eqref{eq:corrEpCosTheta}. 
This shows that our proposed experimental configuration can maintain sufficient phase coherence to perform a Bell test. 
Specifically, the halo width and the coherence length and time are adequate to preserve the necessary interference effects. 

\figThree

The experimental implementation will use a microchannel plate and delay line detector positioned in the far field to detect individual atoms with high spatial and temporal resolution \cite{manning2010HanburyBrownTwissEffect}. 
The dual-species interferometer spans $\sim 100 \mu \mathrm{m}$ in length, while atoms will fall approximately 0.85m to reach the detector, allowing for measurement in the far field with single atom 3D resolution \cite{thomas2022MatterwaveRarityTapsterInterferometer, shin2019BellCorrelationsSpatially}. 
These far-field positions can be mapped to corresponding momentum states after the interferometer sequence (Eqn.~\eqref{eq:psi-t2-uu}-\eqref{eq:psi-t2-dd}).
We can apply a magnetic field gradient during free fall to spatially separate the two species \cite{shin2019BellCorrelationsSpatially,thomas2022MatterwaveRarityTapsterInterferometer}.

Each experimental run would implement independently selected mixing settings $(\varphi_\mA, \varphi_\mB)$ addressing the $\mA$ and $\mB$  interferometer arms, respectively. 
Within each run every azimuthal sector $\phi$ realises a simultaneous copy of the interferometers that shares the same  $(\varphi_\mA, \varphi_\mB)$, thereby greatly increasing the data acquisition rate. 
The CHSH correlator $\mathcal S$ (Eqn~\eqref{eq:chsh-param-def}) is measured experimentally from run-by-run coincidence counts of single-atom detection events, conditioned on the chosen $(\varphi_\mA, \varphi_\mB)$ as in Eqns~\eqref{eq:prob-corr}-\eqref{eq:e-corr-def}.  Single-atom 3D momentum-resolved detection is achieved for \tHeM atoms using a multi channel plate and delay-line detector (MCP-DLD) \cite{manning2010HanburyBrownTwissEffect}.

In principle, one can enforce spacelike separation between the setting-outcome pairs $(\varphi_\mA, \{ \ket{\mA \ULA}, \ket{\mA \DLA} \})$ and $(\varphi_\mB, \{ \ket{\mB \URA}, \ket{\mB \DRA} \})$, 
    by discarding the cylindrical symmetry and addressing only two diametrically opposed momentum modes of each species, i.e. using only the labelled modes in \figurename~\ref{fig:all-in-one}.(b)-(d).  This defines two spatially separated sides of the experiment on the left $(\ULA, \DLA)$ and right $(\URA, \DRA)$ modes, respectively.
Independent random number generators on the two sides could select $\varphi_\mA$ and $\varphi_\mB$, which are applied at $t_2$ before a binary outcome (up or down) is detected from each pair of two momentum modes.
As long as the pair creation, phase setting and detection events for each side are kept outside each other’s light cones, spacelike separation would be achieved, although it would be technically very challenging to implement. 
However, it is important to emphasise that the intention of this proposal is not for a complete loophole-free Bell test, and to note that a loophole-free test took over 30 years \cite{hensen2015LoopholefreeBellInequality} to achieve with photons following the first experimental CHSH Bell demonstration \cite{aspect1981ExperimentalTestsRealistic}.  Rather, this proposal seeks to extend Bell tests to an entirely new type of system, in this case entanglement between atoms of different mass and momentum.

There are several possibly deleterious effects that would be present in an experimental realisation of our proposal, which are not accounted for in our numerical simulation.  Such effects could reduce the amplitude of the measured $\mathcal S$ parameter, to the point where it may no longer be possible to prove non-locality (which requires $\mathcal S > 2$).  The finite mode-occupancy $n$ of experimental halos \cite{hodgman2017SolvingQuantumManyBody} and non-zero detector resolution \cite{manning2010HanburyBrownTwissEffect} could lead to some non-entangled pairs being counted, reducing the measured value of $\mathcal S$.  However, our previous experiments \cite{shin2019BellCorrelationsSpatially,athreya2025bellcorrelationsmomentumentangledpairs} have shown that these issues can be largely circumvented by operating in the low mode-occupancy regime and choosing appropriate bin sizes for the MCP-DLD (Multi channel plate and delay-line detector).  It should also be noted that the quantum efficiency of the detection system alone will not limit $\mathcal S$, due to only using correlated pairs, although it could worsen effects due to finite $n$.  

There are also other mechanisms that would be harder to predict, such as imperfect Bragg pulses or stray magnetic field gradients.  The latter would cause the trajectories taken by the different mass components of the entangled state to diverge due to their different masses, leading to decoherence of the entangled pairs, which would reduce $\mathcal S$.  However, we note that such decoherence effects did not prevent our previous measurements of Bell correlations in spin \cite{shin2019BellCorrelationsSpatially} and momentum \cite{athreya2025bellcorrelationsmomentumentangledpairs}, and similar levels of decoherence and degradation of $\mathcal S$ could be expected here, since many of the relevant factors should be approximately the same across experiments.  In our recent demonstration of momentum entanglement \cite{athreya2025bellcorrelationsmomentumentangledpairs}, we were able to measure a fitted cosine amplitude for an equivalent Bell correlator to $E(\theta_\mA,\theta_\mB)$ of $A_{m}=0.86(3)$.  Assuming similar sources, we would expect a worst case reduction in amplitude of a similar amount here.  Multiplying our fitted cosine amplitude $A=0.865(4)$ (extracted from the purple line $A\cos(\theta_\mA+\theta_\mB)$ in \figurename~\ref{fig:results}) by $A_{m}$ gives $A_{exp}=0.74(3)$, which would reduce our CHSH parameter from $\mathcal S=2.46$ to $\mathcal S_{exp} = 2.12(7)$, which is still greater than the classical value of 2.  This estimates a worst case scenario, as some of the factors that reduce $\mathcal S$ are already accounted for in our simulation, as is shown by the fact that our simulation $A<1$, so in reality the reduction would likely be somewhat less.

\section{Conclusion and Outlook}\label{sec: conclusion and outlook}
In conclusion, we have presented a proposal to test a Bell inequality using momentum-entangled pairs of atoms of different masses.
The distinct atomic transition frequencies allow independent control of the interferometer phases $\varphi$ and pulse durations $\theta$ for each species, 
    enabling a CHSH-type Bell test that overcomes limitations of previous single-species approaches \cite{thomas2022MatterwaveRarityTapsterInterferometer,dussarrat2017TwoParticleFourModeInterferometer}. While our proposal considers \tHeT and \tHeF atoms, the method should in principle be able to be extended to other combinations of species of cold atoms with only minor modifications.

Beyond the Bell test, this setup could also test the Weak Equivalence Principle (WEP) using entangled test masses \cite{geiger2018ProposalQuantumTest}. 
In contrast to the proposal of \cite{geiger2018ProposalQuantumTest}, our scheme could achieve independent time settings in the WEP phase shift parameter 
\begin{align}
\Delta \phi_\mathrm{WEP} = g_\mathcal{A} \frac{|\bmp^\mathcal{A}|}{\hbar} T_\mathcal{A}^2 + g_\B \frac{|\bmp^\B|}{\hbar} T_\B^2,
\end{align}
where $g_{\A,\B}$ are the gravitational accelerations experienced by \thet and \thef, 
while $T_{\A,\B}$ are the times from scattering to mirror pulse for each species.
If our experiment operated with $T=50\mathrm{ms}$ for $N=10^4$ experimental runs, our gravitational sensitivity would be $1/(k T^2 \sqrt{N})\simeq5\codt10^{-7}$, which translates to an E\"otv\"os parameter accuracy of $\eta = \frac{g_\mA-g_\mB}{g_\mA+g_\mB} \simeq 5\cdot10^{-8}$. This sensitivity could also be improved by increasing the fall time via an apparatus that allows long fall times \cite{Hardman_long_drop_2016} or a drop tower \cite{Muentinga_drop_tower_2013}. Extending to $T=1\mathrm{s}$ would give $\eta \simeq 10^{-10}$. While these parameters do not challenge state-of-the-art WEP measurements, where earth based tests achieve $\eta < 10^{-13}$ \cite{Schlamminger_earth_WEP_test_2008} and space based tests an order of magnitude better \cite{Touboul_2022}, the novelty of using a unique and highly quantum test state would still make this an interesting experiment \cite{geiger2018ProposalQuantumTest}.

Furthermore, this setup can be configured to measure Hong-Ou-Mandel (HOM) interference \cite{hong1987MeasurementSubpicosecondTime,LewisSwan2014,Lopes2015}, by using two entangled pairs that are scattered in the same direction (Eqn~\eqref{eq:Psi-total-state-HOM} as the initial state). 
Unlike the photonic case, here both the bosonic ``HOM dip'' (bunching) and fermionic ``HOM peak'' (anti-bunching) could be observed in the same experiment.

There is also the potential to advance our understanding of the interaction between gravity and the entangled states of objects with different masses, such as described in Eqn.~\eqref{eq:initial-state} \cite{carney2019TabletopExperimentsQuantum}. 
According to quantum interpretations, the momentum (or position) of these atoms is fundamentally indeterminate until interacted with, and hence the spacetime metric influenced by their mass is also undetermined \cite{carlesso2019TestingGravitationalField}.  
Note that we are not directly probing gravitationally mediated entanglement or quantised gravitational interactions like gravitons \cite{marletto2017GravitationallyInducedEntanglement}. 
Since general relativity is a local theory, a Bell's inequality violation in an experiment like our proposal would imply that non-local correlations exist between the momentum states of massive particles, which themselves act as sources of spacetime curvature. Any candidate theory of quantum gravity would need to accommodate such observations \cite{ziaeepour2022ComparingQuantumGravity,carlip2008QuantumGravityNecessary}. 
An example of such a theory is gravity-induced collapse models (Di\'osi-Penrose and its similar variants) \cite{diosi_models_1989,penrose_gravitys_1996,penrose_gravitization_2014,bassi_models_2013}, where the collapse rate scales with the difference in the superposed mass distributions. Our momentum entangled state places two different masses into spatially separated momentum superpositions.  Since the entanglement is in external degrees of freedom, it cannot be factored out when writing the entangled state.  While our proposal would not directly demonstrate entanglement of mass (as mixing different mass states is experimentally unfeasible), a Bell test in momentum on such a state would still enable potentially tighter bounds than any photon based or internal-state Bell tests for such collapse models. Other possible theories that could be tested with this state include time-dilation-induced dephasing models  \cite{pikovski_universal_2015,paiva_non-inertial_2022}, and quantum reference frame theories \cite{giacomini2023einsteinsequivalenceprinciplesuperpositions}.
This lays the groundwork for a new class of experiments that explore the boundaries between quantum mechanics and gravity.

\begin{acknowledgements}
The authors would like to thank S.A. Haine for careful reading of the manuscript 
    and J.J. Hope, K.F. Thomas and B.M. Henson for helpful discussions. 
This work was supported through the Australian Research Council (ARC) Discovery Projects, Grant No. DP190103021, DP240101346 and DP240101441. 
S.S. Hodgman was supported by the Australian Research Council Future Fellowship Grant No. FT220100670.
S. Kannan was supported by an Australian Government Research Training Program scholarship. 
\end{acknowledgements}
\printbibliography

\end{document}